\documentclass[english,aps,prl,twocolumn,preprintnumbers,nofootinbib,floatfix]{revtex4-1}
\usepackage{graphicx}

\begin{document}

\title{Graviton mass might reduce tension between early and late time cosmological data}

\date{\today}

\author{Antonio \surname{De Felice}}
\affiliation{Center for Gravitational Physics, Yukawa Institute for Theoretical Physics,
 Kyoto University, 606-8502, Kyoto, Japan}
\author{Shinji Mukohyama}
\affiliation{Center for Gravitational Physics, Yukawa Institute for Theoretical Physics,
 Kyoto University, 606-8502, Kyoto, Japan}
 \affiliation{Kavli Institute for the Physics and Mathematics of the Universe (WPI), UTIAS, The University of Tokyo, 277-8583, Chiba, Japan} 

\preprint{YITP-16-86, IPMU16-0096}

 \begin{abstract}
  The standard $\Lambda$-CDM predicts a growth of structures which tends to be higher than the values of redshift space distortion (RSD) measurements, if the cosmological parameters are fixed by the CMB data. In this paper we point out that this discrepancy can be resolved/understood if we assume that the graviton has a small but non-zero mass. In the context of the Minimal Theory of Massive Gravity (MTMG), due to infrared Lorentz violations measurable only at present cosmological scales, the graviton acquires a mass without being haunted by unwanted extra degrees of freedom. While the so-called self-accelerating branch of cosmological solutions in MTMG has the same phenomenology for the background as well as the scalar- and vector-type linear perturbations as the $\Lambda$-CDM in General Relativity (GR), it is possible to choose another branch so that the background is the same as that in GR but the evolution of matter perturbations gets modified by the graviton mass. On studying the fit of such modified dynamics to the above-mentioned RSD measurements, we find that the $\Lambda$-CDM model is less probable than MTMG by two orders of magnitude. With the help of the cross correlation between the integrated Sachs-Wolfe (ISW) effect and the large scale structure (LSS), the data also pin-down the graviton mass squared around $\mu^2\approx - (3\times 10^{-33} \rm{eV})^2$, which is consistent with the latest bound $|\mu^2|<(1.2\times 10^{-22} \rm{eV})^2$ set by the recent LIGO observation. 
 \end{abstract}

\maketitle

Most recent low-redshift (i.e.\ late-times) cosmological data~\cite{z0.02,z0.067,z0.15,z0.17,z0.17-0.35-0.77,z0.22-0.41-0.6-0.78,z0.25-0.37,z0.3-0.4-0.5-0.6,z0.32-0.57,z0.35,z0.44,z0.59,z0.6-0.73,z0.77,z0.8,z1.36} describing the growth of structures tend to be in tension with respect to high-redshift (i.e.\ early-times) CMB data. The $\Lambda$-CDM in GR is in excellent agreement with data from CMB experiments such as Planck~\cite{Planck}.  Nonetheless, once the cosmological parameters are fixed by the CMB data, GR predicts a growth of structures which tends to be higher than the values of RSD measurements. When the perihelion shift of Mercury was found in the 19th Century, people tried to explain it by introducing an unknown planet called ``Vulcan,'' so to speak a dark planet. The right answer, however, was not a dark planet but to change gravity, from Newton's theory to GR. With this in mind, while the above mentioned discrepancy between the early and late time cosmological data might actually be a result of unknown systematic errors that could come from baryon physics, mass functions, etc., it is definitely intriguing to ask whether the tension can be relaxed by modifying gravity at long distances. In GR, because of gravity's attractive nature, adding dynamical dark energy usually enhances the growth of structures and thus does not seem to help. In fact, history might actually repeat itself. Masses and spins are the most fundamental properties of particles and fields. For this reason, one of the most interesting possibilities for modification of gravity is to give a mass to the graviton, a spin-$2$ particle mediating gravity.  The purpose of the present paper is to point out that the discrepancy between early and late time cosmological data can be resolved/understood if we assume that the graviton has a small but non-zero mass.

While a nonlinear theory of massive gravity stable around a Minkowski background, called the dRGT theory, was discovered in 2010~\cite{dRGT}, it was later shown that all homogeneous and isotropic cosmological solutions in the theory are unstable~\cite{instability}. The MTMG~\cite{MTMG} has then been introduced in order to get rid of the unwanted,  unstable degrees of freedom.  By explicitly breaking Lorentz invariance at cosmological scales, constraints were imposed to the system as to leave only the tensor modes to propagate, as in GR. The MTMG was originally formulated in the Hamiltonian formalism in \cite{MTMG} and then its action was obtained in \cite{MTMG2}. Adopting the ADM decomposition of the $4$-dimensional physical metric, $g_{\mu\nu}dx^{\mu}dx^{\nu}=-N^2dt^2+\gamma_{ij}(dx^i+N^idt)(dx^j+N^jdt)$, and the fiducial metric, $f_{\mu\nu}dx^{\mu}dx^{\nu}=-M^2+\tilde{\gamma}_{ij}(dx^i+M^idt)(dx^j+M^jdt)$, the MTMG action in the unitary gauge is of the form 
\begin{eqnarray}
 S & = & \frac{1}{16\pi G_N}\int d^4x\sqrt{-g}\left\{ R
  +m^2\sum_{n=1}^4 c_nL_{n}
  \right. \nonumber\\
 & & \left.
  + \left(m^{2}\lambda\,\frac{M}{N}\right)^2 F
  + m^2\frac{M}{N}\left[\lambda\tilde{\mathcal{C}}_{0}-(\mathcal{D}_j\lambda^{i})\tilde{\mathcal{C}}^j{}_{i}\right]\right\}\,,
\end{eqnarray}
where $G_N$ is Newton's constant, $R$ is the Ricci scalar of $g_{\mu\nu}$, $m$ is a mass parameter, $c_n$ ($n=1,\cdots,4$) are dimensionless free parameters, $L_n$ are scalars made of ($\gamma_{ij}$, $\tilde{\gamma}_{ij}$, $M/N$), $F$ is a scalar made of ($\gamma_{ij}$, $\tilde{\gamma}_{ij}$) depending bilinearly on $c_n$, $\lambda$ and $\lambda^i$ are auxiliary fields that behave as a scalar and a spatial vector, $\mathcal{D}_i$ is the spatial covariant derivative compatible with $\gamma_{ij}$, and $\tilde{\mathcal{C}}_0$ and $\tilde{\mathcal{C}}^j{}_i$ ($i,j=1,2,3$) are a scalar and a spatial tensor made of ($\gamma_{ij}$, $\tilde{\gamma}_{ij}$, $K_{ij}$) depending linearly on $c_n$~\cite{MTMG2}. Here, $K_{ij}=(\partial_t\gamma_{ij}-\mathcal{D}_iN_j-\mathcal{D}_jN_i)/(2N)$ is the extrinsic curvature and $N_i=\gamma_{ij}N^j$. Since $\tilde{\mathcal{C}}_0$ depends on $K_{ij}$, the kinetic structure of the theory after integrating out $\lambda$ differs from GR and dRGT, as opposed to any other massive gravity theories considered before. The action depends on $m$ and $c_n$ only through the combinations $m^2c_n$. We suppose that we rescale $m$ so that  $c_n=O(1)$ and that $m$ sets the overall scale of the modification of gravity.

There are two distinct branches of cosmological solutions in MTMG: one branch (the self-accelerating branch) that provides a stable nonlinear completion of the self-accelerating solution~\cite{selfacceleration} in dRGT theory and the other branch that we shall consider in the present paper. In the latter branch, often called the normal branch, of MTMG, it is possible to choose the fiducial metric~\footnote{The fiducial metric, in unitary gauge, corresponds to a given external symmetric (0,2) tensor field. In the following we will consider this external field to be homogeneous and isotropic, but time-dependent, in order for FLRW solutions to exist. Still, we have the freedom of choosing its lapse and scale-factor functions. Taking the advantage of this freedom, we set  the ratio of the scale factor of the fiducial metric to that of the physical metric to be a constant $X_0$.} so that the cosmological background behaves exactly as the $\Lambda$-CDM. Therefore, on the background, MTMG is described by one single free parameter $\rho_{\Lambda}$ stemming from the graviton mass~\footnote{\label{footnote:rhoLambda}In terms of the constant ratio $X_0$, we have $\rho_{\Lambda}\equiv m^{2} \,(c_{1}X_{0}^{3}+3c_{2}X_{0}^{2}+3c_{3}X_{0}+c_{4})/(16\pi G_N)$.}, as
\begin{equation}
 3H^{2}  = 8\pi G_N\, (\rho_{m}+\rho_{\Lambda})\,,\
  (H^2)' =  -8\pi G_N\,(\rho_{m}+P_{m})\,. \label{eqn:backgroundEOM}
\end{equation}
where $H$ is the Hubble expansion rate, $\rho_{m}$ and $P_{m}$ are energy density and pressure of matter, and a prime denotes derivative with respect to the e-fold time-variable: $\mathcal{N}=-\ln(1+z)$, and $z$ is the cosmological redshift.
Furthermore, on studying the influence of the theory on the matter sector, it was found that in cosmological linear perturbation theory, the evolution of each mode differs from the $\Lambda$-CDM only for low redshifts \cite{MTMG2}. In fact, by choosing the time variable as $t=\mathcal{N}$ and the lapse function as $N=1/H$ in the equation of motion for the dust perturbation $\delta_m$ in the matter comoving gauge that was derived in \cite{MTMG2}, one sees that the dynamics of dust perturbations is described by
\begin{equation}
 \delta_{m}''+\left[2C_5(\mathcal{N},k^2) + \frac{(H^2)'}{2H^2}\right]\delta_{m}'
  - \frac{4\pi\rho_{m}}{H^2}G_{\rm eff}(\mathcal{N},k^2)\delta_{m} = 0\,,
\end{equation}
where $C_5(\mathcal{N},k^2)$ and $G_{\rm eff}(\mathcal{N},k^2)$ are functions of $\mathcal{N}$ and the squared comoving wavenumber $k^2$, and depend also on $c_n$ ($n=1,\cdots,4$). It is notable that in MTMG coupled with the CDM dust fluid the number of physical degrees freedom in the scalar sector is one, corresponding to the dust perturbation, and there is no extra degree of freedom. For this reason we do not need to rely on the quasistatic approximation, that is commonly adopted in many modified gravity theories. By using the background equations (\ref{eqn:backgroundEOM}) with $P_m=0$, it is shown that $C_5=1+O((aH)^2/k^2)$ and $G_{\rm eff}=\bar{G}_{\rm eff}+O((aH)^2/k^2)$, where 
\begin{equation}
 \bar{G}_{\rm eff} =
  \frac{2}{3}G_{N}
  \left[\frac{3}{2-\theta Y}-\frac{9\theta Y\Omega_{m}}{2(\theta Y-2)^{2}}\right]\,, 
  \label{eqn:Geff}
\end{equation}
$\Omega_{m}  = 8\pi\,G_N\rho_{m}/(3H^{2})$, $Y \equiv H_{0}^{2}/H^{2}$, $\theta \equiv \mu^2/H_0^2$, $H_0=H(\mathcal{N}=0)$, and $\mu$ is the mass of the gravitational waves~\footnote{In terms of $X_0$, we have $\mu^2=m^{2}X_{0}(c_{1}X_{0}^{2}+2c_{2}X_{0}+c_{3})/2$ (see footnote \ref{footnote:rhoLambda} as well).}. On taking the subhorizon limit $k^2\gg H^2$ but without need for the quasistatic approximation, one thus obtains a simple equation without explicit dependence on $k$ as
\begin{equation}
 \delta_{m}''+\left(2-\frac{3}{2}\,\Omega_{m}\right)\delta_{m}'
  -\frac{3}{2}\frac{\bar{G}_{\rm eff}}{G_N}\Omega_{m}\delta_{m}=0\,.\label{eq:pert}
\end{equation}
Therefore, at the level of linear perturbations in the subhorizon limit~\footnote{The dependence on other combinations of $c_n$ ($n=1,\cdots,4$) shows up at the order $O((aH)^2/k^2)$ and in nonlinear corrections.}, we find the second free parameter, $\theta$. In order for the tensor modes not to develop instability whose time scale is shorter than the age of the universe~\footnote{The precise value of the lower bound on $\theta$ does not change the result since the likelihood quickly decreases for smaller $\theta$.}, we require that $\theta\geq -10$. In the limit $\theta\to0$, we recover the evolution equation for the perturbations in the $\Lambda$-CDM. This same limit is achieved by $Y\to0$, i.e.\ at early times. The evolution equations for $\Omega_{m}$ and $Y$ read as follows 
\begin{equation}
\Omega_{m}' = 3\Omega_{m}(\Omega_{m}-1)\,,\ Y' = 3Y\Omega_{m}\,.
\end{equation}
Furthermore we impose the system to satisfy the following boundary conditions at a point of high redshift, e.g.\ $\mathcal{N}=\mathcal{N}_{i}=-6$ corresponding to $z=z_i\simeq 402.4$, and at the present time, $\mathcal{N}=0$: $\delta_{m}'(\mathcal{N}_{i}) = \delta_{m}(\mathcal{N}_{i})$ (selecting the growing mode at early times), $Y(\mathcal{N}=0) = 1$ (by definition), and $\Omega_{m}(\mathcal{N}=0) = 0.3089$ (fixing, once for all, the only background parameter $\rho_{\Lambda}$ to the $\Lambda$-CDM best-fit value\cite{Planck}; as we shall see later, the ISW effect due to the time-dependence of $\bar{G}_{\rm eff}$ does not change the $\Lambda$-CDM best-fit value significantly for a suitable choice of $\theta$). One may set $\delta_{m}(\mathcal{N}_{i})$ to any non-zero value since the overall amplitude of $\delta_m$ does not affect the observable defined below.

The observable we will use to constrain the only remaining free
parameter, $\theta$, is defined as $y(z)\equiv f(z)\sigma_{8}(z)$,
where $f(z)=\delta_{m}'/\delta_{m}$, and $\sigma_{8}(z)$ is
the rms mass fluctuation of a sphere of radius 8 Mpc. Assuming a
window function which is only dependent on $k$ and on the radius of
the spherical distribution of mass, we find that
$\sigma_{8}(z)\propto\delta_{m}(z)$. We can thus write
$\sigma_{8}(z)=\sigma_{8}(z_i)\,\delta_{m}(\mathcal{N})/\delta_{m}(\mathcal{N}_{i})$. 
Since GR and MTMG are indistinguishable at early times and their backgrounds are exactly the same for all times, the CMB data give the same constraint on the value(s) of $\sigma_{8}(z_i)$ (and $\Omega_{m}(\mathcal{N}=0)$) for both theories. In GR we know that its best fit is given by $\sigma_{8}^{\rm GR}(z=0)=0.8159$ and we can determine $\sigma_{8}(z_i)$ (common for GR and MTMG) by using the evolution for $\delta_{m}^{{\rm GR}}$, so that $\sigma_{8}(z_i)=0.8159\,\delta_{m}^{{\rm GR}}(\mathcal{N}_{i})/\delta_{m}^{{\rm GR}}(0)$.
This prescription is justified since the growth function in MTMG on sub-horizon scales is scale-independent. Having defined the observable, $y(z)$, we can now introduce the chi-square functions as follows
\begin{equation}
 \chi_{{\rm MTMG}}^{2}\equiv\sum_{n}\frac{\bigl(y_{n}-y_{n}^{{\rm MTMG}}\bigr)^{2}}{\sigma_{n}^{2}}\,,\ 
  \chi_{{\rm GR}}^{2}\equiv\sum_{n}\frac{\bigl(y_{n}-y_{n}^{{\rm GR}}\bigr)^{2}}{\sigma_{n}^{2}}\,,
\end{equation}
where the index $n$ runs over the data points reported in Table 1, $y_n$ and $\sigma_n$ are the observed values of $y$ and its uncertainty for the $n$-th data point, and $y_{n}^{{\rm MTMG}}$ and $y_{n}^{{\rm GR}}$ are the corresponding theoretical predictions in MTMG and GR, respectively. Notice that $\chi_{{\rm GR}}^{2}$ is not a function of any parameter, but merely a number. On the other hand, $\chi_{{\rm MTMG}}^{2}$ is a function only of the parameter $\theta$ (all the initial conditions are completely fixed).
\begin{table}[ht]
  \caption{Data points}
  \begin{center}
\begin{tabular}{l | l | l | l}
$z$ & $\mathcal{N}$ & $f\sigma_8$ & Refs. \\
\hline
0.02  & -0.020 & 0.360  $\pm$  0.04 & \cite{z0.02} \\
0.067  & -0.065 & 0.423  $\pm$  0.055 & \cite{z0.067} \\
0.15  & -0.14 & 0.490  $\pm$  0.15 & \cite{z0.15} \\
0.17  & -0.16 & 0.510  $\pm$  0.06 & \cite{z0.17,z0.17-0.35-0.77} \\
0.22  & -0.20 & 0.420  $\pm$  0.07 & \cite{z0.22-0.41-0.6-0.78} \\
0.25  & -0.22 & 0.351  $\pm$  0.058 & \cite{z0.25-0.37} \\
0.3  & -0.26 & 0.408  $\pm$  0.0552 & \cite{z0.3-0.4-0.5-0.6} \\
0.32  & -0.28 & 0.394  $\pm$  0.062 & \cite{z0.32-0.57} \\
0.35  & -0.30 & 0.440  $\pm$  0.05 &  \cite{z0.17-0.35-0.77,z0.35} \\
0.37  & -0.31 & 0.460  $\pm$  0.038 & \cite{z0.25-0.37} \\
0.4  & -0.336 & 0.419  $\pm$  0.041 &  \cite{z0.3-0.4-0.5-0.6} \\
0.41  & -0.34 & 0.450  $\pm$  0.04 &  \cite{z0.22-0.41-0.6-0.78} \\
0.44  & -0.36 & 0.413  $\pm$  0.08 &  \cite{z0.44} \\
0.5  & -0.41 & 0.427  $\pm$  0.043 &  \cite{z0.3-0.4-0.5-0.6} \\
0.57  & -0.45 & 0.444  $\pm$  0.038 & \cite{z0.32-0.57} \\
0.59  & -0.46 & 0.488  $\pm$  0.06 & \cite{z0.59} \\
0.6  & -0.47 & 0.430  $\pm$  0.04 & \cite{z0.22-0.41-0.6-0.78} \\
0.6 & -0.47 & 0.390    $\pm$  0.063 & \cite{z0.6-0.73} \\
0.73  & -0.55 & 0.437  $\pm$  0.072 & \cite{z0.6-0.73} \\
0.77  & -0.57 & 0.490  $\pm$  0.18 &  \cite{z0.77,z0.17-0.35-0.77} \\
0.78  & -0.58 & 0.380  $\pm$  0.04 &  \cite{z0.22-0.41-0.6-0.78} \\
0.8  & -0.59 & 0.470 $\pm$  0.08 &  \cite{z0.8} \\
1.36  & -0.86& 0.482  $\pm$  0.116 & \cite{z1.36} \\
\end{tabular}
\end{center}
\end{table}
On using the data points reported in Table 1, we plot $\chi_{{\rm MTMG}}^{2}(\theta)$ in Fig.\ 1. It should be noticed that for non-negative $\theta$, $\chi_{{\rm MTMG}}^{2}(\theta)$ has a local minimum at $\theta_{\rm min}\approx1.165$ and then rapidly increases for larger values of $\theta$. A similar behavior is observed for negative $\theta$, with another local minimum at $\theta_{\rm min}\approx -3.828$.  

 \begin{figure}[ht]
%\centering
 \includegraphics[width=0.5\textwidth]{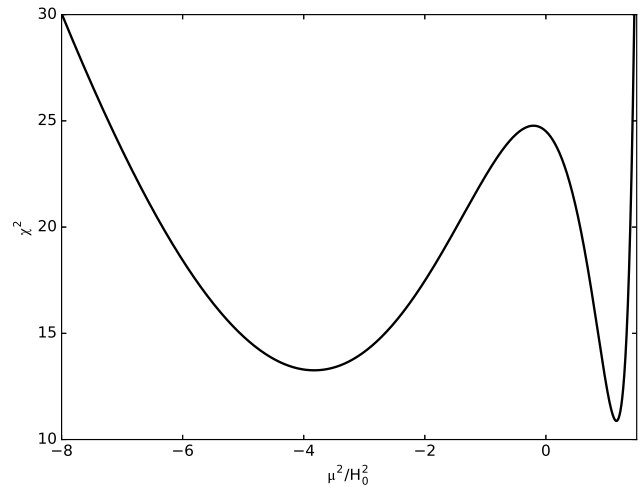}
\caption{The chi-square for MTMG, $\chi_{{\rm MTMG}}^{2}$, as a function of the parameter $\theta=\mu^2/H_0^2$. GR is recovered for $\theta=0$.}
%\end{center}
 \end{figure}

 In order to discriminate between the two local minima of $\chi_{{\rm MTMG}}^{2}(\theta)$, we now consider the ISW effect. The decrease of $\bar{G}_{\rm eff}$ at late-time, indicated by the formula (\ref{eqn:Geff}) and shown in Fig.\ 4, can result in some additional ISW contribution to the CMB anisotropies since the ISW contribution to the anisotropies is a weighted line-of-sight integral of $\Phi'+\Psi'$, which depends on $\bar{G}_{\rm eff}$. Unlike the case of scalar-tensor theories~\cite{Kimura:2011td}, the ISW effect and the LSS can either correlate or anti-correlate, depending on the sign of $\theta$. This is because MTMG does not have an extra degree of freedom, while scalar-tensor theories do.  For $\theta\approx1.165$, adopting the large $k$ approximation, the correlation between the ISW effect and the LSS is shown to be negative. This means that $\theta\approx1.165$ is ruled out by observational data (see e.g.\ Fig 1 of \cite{Giannantonio:2012aa}). On the other hand, for $\theta\approx -3.828$, the correlation is positive as in the observational data (as well as in the $\Lambda$-CDM). Since MTMG studied in the present paper has the same background evolution as the $\Lambda$-CDM and the error bars for the ISW-LSS correlation data are large, this implies that CMB observations fix $\Omega_{m}$ to essentially the same value in MTMG with $\theta\approx -3.828$ and the $\Lambda$-CDM.

Going back to the observable $y(z)\equiv f(z)\sigma_{8}(z)$, on
defining the likelihood function as $\mathcal{L}=\exp[-\chi_{{\rm
      MTMG}}^{2}/2]$, and sampling it via the MCMC method\footnote{For
  this aim, we have made use of the \texttt{emcee} package
  \cite{emcee}.}, we obtain, in Fig.\ 2, a likelihood plot for the
free parameter $\theta$, leading to
$\theta=-3.828^{{+}0.875}_{{-}0.962}$ at 68.27\% C.L.\footnote{We find
  that, for the alllowed best-fit ($\theta\approx-3.828$),
  $\sigma_8(0)\approx0.795$.}

 \begin{figure}[ht]
  \begin{center}
 \includegraphics[width=0.5\textwidth]{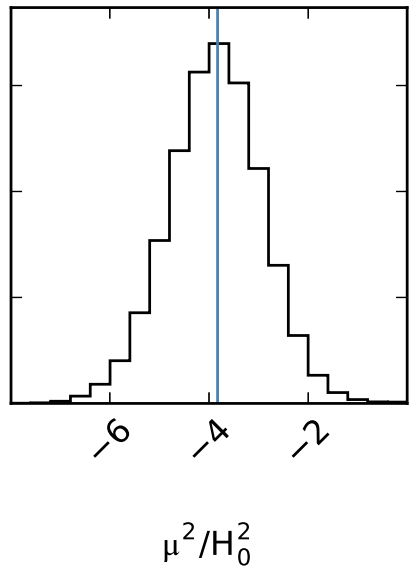}
  \caption{Distribution of the parameter $\theta=\mu^2/H_0^2$ near $\theta\approx -3.828$, according to the likelihood function defined via the $\chi_{{\rm MTMG}}^{2}$. We have given flat prior to the parameter $\theta$, in the range $-10\leq\theta\leq0$. Smaller negative values for $\theta$ lead to a bad fit to the data, whereas positive values of $\theta$, either have a worse fit, or lead to anti-correlations to the ISW-effect, or, for even larger values, lead $Y$ to reach the point $Y_{\infty}=2/\theta$, at which $\bar{G}_{\rm eff}$ switches sign and the last term in (\ref{eq:pert}) diverges. The blue line indicates the value of the maximum likelihood point, i.e.\ the minimum of $\chi_{{\rm MTMG}}^{2}$.}
\end{center}
 \end{figure}

Besides, we numerically find that, for MTMG, $\chi^{2}_{\rm MTMG}(\theta)$ 
possesses local minima as $\chi_{{\rm MTMG}}^{2}(\theta_{\rm min}\approx -3.828)\equiv\bar\chi^2_{\rm MTMG}\approx 13.259$ whereas, for GR, we have $\chi_{{\rm GR}}^{2}=\chi_{{\rm MTMG}}^{2}(\theta=0)\approx24.51$. 

Following the Akaike Information Criterion (AIC)~\cite{AIC}~\footnote{In the absence of an established fundamental principle to determine the prior probability distribution in the space of theories including gravity, we avoid detailed Baysian analysis.}, used to compare the relative likelihood of two models, we find that in this case GR is $\exp[(\bar\chi^2_{\rm MTMG}+2n_{\rm fit}-\chi^2_{\rm GR})/2]\approx1\times10^{-2}$ as probable as MTMG, where $n_{\rm fit}=1$ is the number of extra fitting parameter(s) in (the subhorizon limit of) MTMG. This result is already interesting in terms of model building, as it states that the data lead to a larger likelihood for MTMG compared to the $\Lambda$-CDM. Moreover, from the theoretical point of view, the RSD measurements do set the value of the graviton mass squared to be $\mu^2=\theta\,H_0^2=-3.828^{{+}0.875}_{{-}0.962}\,H_0^2$. This is consistent with the bound on the graviton mass set by the LIGO collaboration~\cite{LIGO}, which applies to $\mu^2$ since it is the mass squared entering the dispersion relation of gravitational waves. Also, this value of the graviton mass squared means that tensor modes at the present horizon scale or longer scales may grow now and in the future. Although such a slow growth is at present difficult to observe, it is certainly interesting to look for its observable signatures in the future. It should be noted that for the best-fit value of $\theta$, we find that $|\mu|\simeq H_0$. This implies that in order to fit the data we do not need to add any new tuning/hierarchy among the physical scales in addition to the scale of the acceleration.

Finally, in Fig.\ 3, we plot the data and the GR fit (red dashed line) together with the best MTMG fit (thick black line), whereas in Fig.\ 4 we show the evolution of the effective gravitational constant for the perturbations $\bar{G}_{\rm eff}/G_{N}$, as a function of redshift and of $1/Y=\rho_{\rm tot}/\rho_{{\rm tot},0}$, where $\rho_{\rm tot}=\rho_{\rm m}+\rho_{\Lambda}$ is the total energy density and $\rho_{{\rm tot},0}$ is its present value.

 \begin{figure}[ht]
%    \centering
 \includegraphics[width=0.5\textwidth]{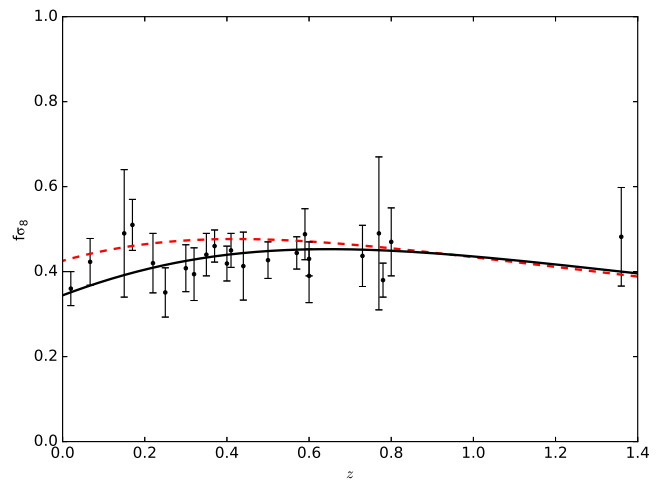}
\caption{Fit to the data for GR (red dashed line), and MTMG (thick black line). For the source of each data point, see Table 1.}
 \end{figure}

 \begin{figure}[ht]
%    \centering
 \includegraphics[width=0.5\textwidth]{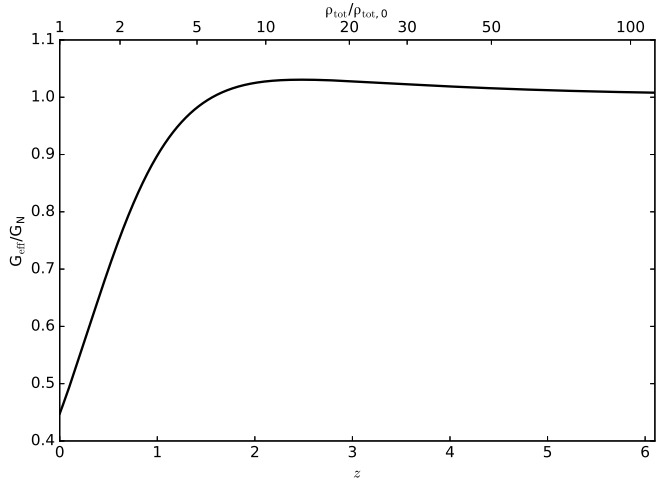}
\caption{Evolution of $\bar{G}_{\rm eff}/G_N$ as a function of redshift (lower $x$-axis), and of 
$\rho_{\rm tot}/\rho_{\rm tot,0}$ (upper $x$-axis), for the best fit of MTMG.}
 \end{figure}

Fig.\ 4 shows deviation of MTMG from GR, $|\bar{G}_{\rm eff}/G_{N}-1|>0.01$, only for $z<5.49$, which translates to $\rho_{\rm tot}<85\rho_{\rm tot,0}$. This observation, combined with the fact that in MTMG there is no scalar/vector degree of freedom to screen \cite{MTMG}, indicates that we will recover GR when the matter density of the environment, $\rho_{\rm env}$, is much higher than $\rho_{\rm tot,0}$. For example, inside the galaxy and the solar system, $\rho_{\rm env}$ is high enough to suppress any deviations from GR. On the other hand, as for the growth of LSS at low redshift, corresponding to low $\rho_{\rm env}$, Figs.\ 3 and 4 clearly show deviations of MTMG from GR, which greatly help reconciling the RSD data to the CMB data.

In summary, in the context of MTMG a small but non-zero graviton mass tends to reduce the tension between early-time and late-time data sets. It also provides a model for the evolution of matter perturbations which can be further studied for the implications that the existence of a non-zero-mass graviton might have. This model can be further tested against future experiments/measurements related to the cosmological-scale dynamics of the CDM dust fluid.

\begin{acknowledgments}
ADF was supported by JSPS KAKENHI Grant Numbers 16K05348,
    16H01099. SM was supported in part by JSPS KAKENHI Grant Number 24540256 and
World Premier International Research Center Initiative (WPI), MEXT,
Japan.
\end{acknowledgments}

\end{document}